\documentclass[10pt,english,a4paper,twoside,twocolumn]{revtex4-1}
\usepackage{amsfonts,hyperref}

\usepackage{graphicx}
\usepackage{amsmath}
\usepackage{amssymb}
\usepackage{bm}% bold math
\usepackage{gensymb}
\usepackage{braket}
\usepackage{mathtools}

\usepackage{algorithm}
\usepackage{algpseudocode}

\newcommand{\floor}[1]{\lfloor #1 \rfloor}

\begin{document}

\title{Simulating the out-of-equilibrium dynamics of local observables by trading entanglement for mixture}

\author{J. Surace}
\affiliation{Department of Physics and SUPA, University of Strathclyde, Glasgow G4 0NG, United Kingdom}
\author{M. Piani}
\affiliation{Department of Physics and SUPA, University of Strathclyde, Glasgow G4 0NG, United Kingdom}
\author{L. Tagliacozzo}
\affiliation{Department of Physics and SUPA, University of Strathclyde, Glasgow G4 0NG, United Kingdom}
\affiliation{Departament de F\'{\i}sica Qu\`antica i Astrof\'{\i}sica and Institut de Ci\`encies del Cosmos (ICCUB), Universitat de Barcelona,  Mart\'{\i} i Franqu\`es 1, 08028 Barcelona, Catalonia, Spain}

\begin{abstract}The fact that the computational cost of simulating a many-body quantum system on a computer increases with the amount of entanglement has been considered as the major bottleneck for simulating its out-of-equilibrium dynamics.
Some aspects of the dynamics are, nevertheless, robust under appropriately devised approximations. Here we present a possible algorithm that allows to systematically approximate  the equilibration value of local operators after a quantum quench.
At the core of our proposal there is the idea to transform entanglement between distant parts of the system into mixture, and at the same time preserving
the local reduced density matrices of the system. 
We benchmark the resulting algorithm by studying quenches of quadratic Fermionic Hamiltonians.
\end{abstract}

\maketitle %% required

\section{Introduction}
%%%%In questa parte prima introduco l'area law e poi DMRG,MPS,TEBD
Simulating the time evolution of many-body quantum systems by classical means is hard.
In fact, simulating an $N$ constituents system requires storage and computation time that scale exponentially with $N$. As an example, consider  the exponentially large space needed in order to store the $2^N$ complex coefficients describing the state of a  $\frac{1}{2}$ spin chain of length $N$.
% This complexity is intrinsic in the formalism and seems that one cannot avoid it if interested in describing a random general quantum state. It can be argued that physics is not about random states in a Hilbert space, but some of them are more relevant then others and thus in order overcome the complexity problem it should be possible to focus on the relevant degrees of freedom to reduce the complexity of the description as statistical mechanics does in  classical physics.
% This is also the general idea behind every compression algorithm and celebrated DMRG algorithm

In some cases the specific structure of the problem allows to devise efficient classical algorithms for simulating many-body systems. For example, in the context of statistical mechanics, a many-body system can have exponentially many configurations. However, in most cases, using Monte-Carlo algorithms, we can sample only a polynomially-large set of those configurations, thus simulating the system efficiently \cite{parisi1988}.

In the context of many-body quantum systems at equilibrium, the locality of correlations \cite{hastings2007,wolf2008,eisert2010}  can be exploited in order to design efficient algorithms based on tensor networks. The exponential decay of correlations in the relevant states at equilibrium, i.e. the ground and thermal states of gapped local Hamiltonians implies that these states only have a limited amount of entanglement \cite{hastings2007,wolf2008}. Their entanglement grows only proportionally to the boundary of a region rather than its volume, a phenomenon formalised in the “area law” of entanglement  \cite{eisert2010,vidal2003, schollwock2011, eisert2013}. States whose entropy fulfils the area law, generally,  can be efficiently described by tensor networks and manipulated by appropriate algorithms. 
The most notable examples are DMRG  \cite{white1993,white1992} and its higher dimensional generalisations (see e.g.  \cite{orus2014,schollwock2011}). DMRG, as of today, provides the most accurate results for strongly correlated one-dimensional quantum systems.

The locality of correlations is lost in the context of many-body quantum systems out-of-equilibrium. In particular it is lost in the dynamics induced by a quench of the Hamiltonian. In this scenario, initially localised correlations can be radiated to arbitrarily large distances \cite{calabrese2005,dechiara2006} leading to a fast growth of entanglement with time \cite{lauchli2008,kim2013,fagotti2015,kormos2017,vonkeyserlingk2018,collura2018}. As a consequence,  traditional tensor network approaches fail to encode the states generated during the  out-of-equilibrium dynamics at relatively short times \cite{trotzky2012}.

%While these observations seem to suggest that simple tensor networks cannot  describe the states emerging at long-time out of equilibrium, they could still be used in order to describe the evolution of  a restricted set of local operators for arbitrary long times. 
Here we focus on encoding locally the states generated during the  out-of-equilibrium dynamic, rather than trying to encode the full states. We design an algorithm aiming at reproducing locally the out-of-equilibrium dynamics for arbitrarily long times.

The algorithm is designed with tensor networks in mind, but here  we start benchmarking it in the context of  non-interacting systems where we can also compute the full out-of-equilibrium dynamics. 

The performances of tensor-network algorithms  mostly depend on the amount of entanglement contained in  the states they try to describe.
The strength of the interactions in a  specific Hamiltonian does not necessarily affect the amount of entanglement between the constituents  in its ground state.
For example, ground states of free systems can be robustly entangled  (think e.g. about a Fermi see in momentum space) while strongly interacting Hamiltonians can have product ground states (think e.g. about the ground state of a Mott insulator). 

The use of non-interacting systems is thus often seen as a first benchmark for tensor network algorithms (see e.g. \cite{corboz2009}). Recently several authors have realised that it is also possible to directly implement tensor network algorithms at the level  of the correlation matrices (see e.g. \cite{evenbly2010a,evenbly2010,fishman2015a,evenbly2016a,haegeman2018}). This discovery allows to prototype new tensor-network ideas without having to fully implement them. It also allows to decouple the effects caused by  the physical approximations contained in the algorithms from those of the spurious approximations  that could appear in specific tensor-networks implementations.

For these reasons here we benchmark  a new tensor network algorithm at the level of correlation matrices  in the case of free-fermions. 
In particular, we approximate the out-of-equilibrium dynamics of a free fermionic system by only using its short-range correlations.

This is only possible by trading entanglement with mixture during the out-of-equilibrium evolution  (similarly to what happens in the algorithm proposed in \cite{white2018,hauschild2017}). As a result, we can approximate  the robustly entangled pure states generated during the evolution with slightly entangled mixed states.

Crucially those mixed states are locally indistinguishable from the pure state they approximate. In this way, we aim at performing the full out-of-equilibrium dynamics in a sub-space of slightly entangled states that we can encode with tensor networks. 

We show that despite, the rough approximations involved in such an algorithm, we are still able to accurately predict the
correct equilibration value of local observables and their approach to equilibrium. This fact relies on the robustness of the  equilibration process 
in spite of ignoring the long-distance correlations generated during the out-of-equilibrium evolution.

\section{Robust aspects of quantum quenches}
In the following we focus on the out-of-equilbirum dynamics after a quantum quench \cite{polkovnikov2011}.
Initially the system is  at equilibrium, say in the ground state $|\psi \rangle$ of some local Hamiltonian $H$. The Hamiltonian is abruptly changed from  $H$ to $\tilde{H}$, quenching the system out-of-equilibrium. The corresponding evolution is described by
\begin{equation}
|\psi(t)\rangle = e^{-it\tilde{H}}|\psi \rangle.
\end{equation}

Cardy and Calabrese  \cite{calabrese2005}
showed that in this setting the entanglement entropy between two different partitions of $|\psi(t)\rangle$ grows linearly in time, a footprint of the radiation of the correlation as pseudo-particles \cite{alba2017,calabrese2005,calabrese2011,calabrese2012,calabrese2012a}. This leads the corresponding states to become too entangled and hard to represent with standard tensor network algorithms after relatively short times \cite{trotzky2012,verstraete2004} \footnote{Here we consider only systems whose excitations are extended and can be described using pseudo particles. A treatment of different systems such as those described in \cite{nahum2017} will be performed separately.}.

The short-time dynamics is highly non-universal and very sensitive to the specific details of the quench \footnote{For a detailed analysis of the universal aspects encoded in the light cone spreading of the correlations after a quench we refer the reader to Ref.  \cite{cevolani2018}.}.
Here we try to address robust features of the out-of-equilibrium dynamics after a quench, that is, features that are not too sensitive to the specific details of the quench.  One of such features is the equilibration of local observables occurring at long times after the quench.

In most cases, the values of the relaxed local observables are indistinguishable from those computed on the  \emph{diagonal ensemble} (DE) defined as
\begin{equation}
\rho_{DE(\tilde{H})}\coloneqq \sum_n |E_n\rangle \langle E_n | \rho |E_n\rangle \langle E_n |,
\end{equation}
where $\{|E_n\rangle\}_n$ are the eigenvectors of the Hamiltonian $\tilde{H}$ driving the dynamics and $\rho$ encodes the state of the system before the quench. That is, if an observable $A$ equilibrates, at late times after the quench  $\langle A(t) \rangle \simeq Tr\left[A\rho_{DE(\tilde{H})}\right]$. The direct construction of the DE is  exponentially hard.

 %the DE can be approximated by an appropriately designed Gibbs ensemble  \cite{deutsch1991,srednicki1994,rigol2008,rigol2012,dalessio2016,rigol2014}. 

 %The fact that the GE is finitely correlated could thus be the reason why our alogorithm works.

 In the generic cases, the DE is locally approximated by a Gibbs Ensemble (GE) as a consequence of the  \emph{eigenstate thermalisation hypothesis} (ETH) \cite{deutsch1991,srednicki1994,rigol2008,rigol2012,dalessio2016,rigol2014}.
%TFor example, the expectation value of local operators at long times after the quench, in a system that thermalises, can be described by a Gibbs state.
The temperature of the Gibbs ensemble only depends on the energy of the initial state unveiling a high degree of robustness in the process of thermalisation.

For those systems described by a local Hamiltonian, the energy is conserved if we conserve short-range correlations. As a result,  by designing an approximate dynamics  conserving short-range correlations,  the robustness of the thermalisation process forces the convergence to the correct state, in spite of discarding the long-range correlations.

The free Fermionic systems we use here as a benchmark do not satisfy the ETH, since
%the Gibbs ensemble is not always the correct local approximation for the relaxed state.
their relaxed state depends on the initial occupation of all the free modes. These occupations are conserved during the evolution.
As a result, the equilibration of these systems is locally described  by \emph{generalised Gibbs ensemble} (GGE) rather than by a GE. The GGE can be defined as the  state that maximizes the entropy at fixed value of all the conserved quantities  \cite{rigol2007,rigol2008,cramer2008,barthel2008,cramer2010,calabrese2012,fagotti2013,calabrese2011, langen2015,ilievski2015,vidmar2016a}.

The equilibrium state after a quench in free Fermionic systems thus depends on infinitely many parameters. This fact seems to  spoil the robustness of the equilibration process. Luckily a weaker notion of robustness can be recovered. The occupation of the free modes can be re-expressed as the conservation of charges whose densities are defined on bounded regions of the lattice. We can thus sort the conserved charges by the dimension of the support of their density.  Charges whose densities have support on smaller blocks are more local than those with support on larger blocks. 

As shown in   Ref. \cite{fagotti2013,essler2017},  the DE is also  locally approximated  by a truncated version of the GDE. The latter only depends on a finite number of parameters associated with conserved charges built out of density with support smaller than a fixed finite number of sites. The accuracy of the approximation increases by including in the truncated GGE charges with larger and larger supports   (see however \cite{pozsgay2014,alba2015,kollar2008,gangardt2008} for cases in which this approximation fails).

 Our benchmarks with free Fermions, thus rely on this weaker notion of robustness.  Any algorithm correctly describing short-range correlations is forced by this weaker robustness to convergence to a truncated version of the GGE. We both design and characterise such an algorithm in the following sections.

%
%

%

%Similar ideas have been pursued in the context of tensor networks \cite{leviatan2017,hauschild2017,white2018} for systems that thermalise.
%Here, by using free Fermionic systems we are able to compare the approximated dynamics to the exact one, and analyse issues such as the physicality of the state obtained during the evolution (that cannot be addressed in the context of tensor networks). In this way we can better understand up to which extent the results presented provide a faithful description of a physical system out-of-equilibrium.

\section{The algorithm}
The algorithm we introduce here is designed to work with tensor-network states. It is inspired by the original  time-evolving block decimation algorithm  (TEBD) \cite{vidal2004a} and the time dependent density matrix renormalization group (t-DMRG) \cite{white2004}.  It exploits both our ability to encode slightly correlated states as tensor networks and  to perform, almost exactly, their short-time dynamics.

Differently from all of these algorithms, it ensures that the approximated state i) has the short-range correlations of the state we want to approximate, ii) can be encoded  with the available resources iii) is a mixed state.

By using i) we exploit the idea of robustness discussed in the previous section; with ii) we ensure that the algorithm is practically useful, also at long-times. The choice of iii) is dictated by the observation that both the GE and GGE are mixed states. 

For completeness, we start by reviewing standard tensor networks techniques for performing the out-of-equilibrium evolution of 1D systems. The main idea was put-forward by G. Vidal in the paper that introduced the TEBD \cite{vidal2004a} and was later refined in several key contributions (see e.g. the recent review on the topic \cite{paeckel2019}).

Given an initial state $\ket{\psi(0)}$ we want to evolve it for a time $t$. In formula  $\ket{\psi(t)}=\exp(-i H t)\ket{\psi(0)}$. Even assuming we know how to encode the initial state (say e.g. it is a product state in the local basis), we still need to apply to it $U(t)=\exp(-i H t)$, that is an  operator that grows exponentially with the size of the system. 

The general strategy can be illustrated by  restricting to local Hamiltonians (see however \cite{koffel2012} for how to extend it to non-local ones), $H=\sum_{\langle i, j \rangle} h_{i,j}$, with $h_{i,j}$ the Hamiltonian density acting on constituents $i,j$ that are nearest neighbours on the lattice. 

We  subdivide the evolution in $M$ small steps  $\exp(-i H t) =\exp(\frac{-i H t}{M})^M=\exp(-i H \delta t)^M$ in such a way that $\delta t =t/M$ becomes arbitrarily small. 
We now have to solve  $M$ short-time evolutions from $t_i \to t_i+\delta t$ where $t_i = (i-1) \delta_t$ and $i=1, \cdots, M$. Each evolution acts on the state that is produced by the previous step.   

%Using the fact that the time-step is small, we can approximate
% $U(\delta t)=\exp(-i H \delta t)$ for example as $U(\delta t)=\mathbb{I}-i H \delta t +{\cal O}(\delta t^2)$. At first order in $\delta t$ in order to perform each of the $M$ short steps we need to solve the following problem, $\ket{\psi(t+\delta_t)}=\left(\mathbb{I}-i H \delta t\right) \ket{\psi(t)}$.
 
 %We can further simplify this task by using the fact that the Hamiltonian is a sum of local terms. Each elementary time step becomes a sum  $\ket{\psi(t_{i+1})}=\sum_{\langle i, j \rangle }\left[\left(\mathbb{I}-i h_{i,j} \delta t\right) \ket{\psi(t_i)}\right]$.
 
In order to proceed we now assume that $\ket{\psi(0)}$ is a matrix product state (MPS), that is a state  of the form $\ket{\psi(0)}=\sum_{i_1 \cdots i_N} c^{i_1 \cdots i_N} \ket{i_1, \cdots i_N}$ with    $c^{i_1 \cdots i_N}= tr \left( A^{i_1}\cdot A^{i_2}\dots A^{i_{N-1}}\cdot A^{i_N}\right)$, where each of the $A^{i}$ is a $D\times D$ matrix.

\begin{figure}[!hbt]
 \includegraphics[width=\columnwidth]{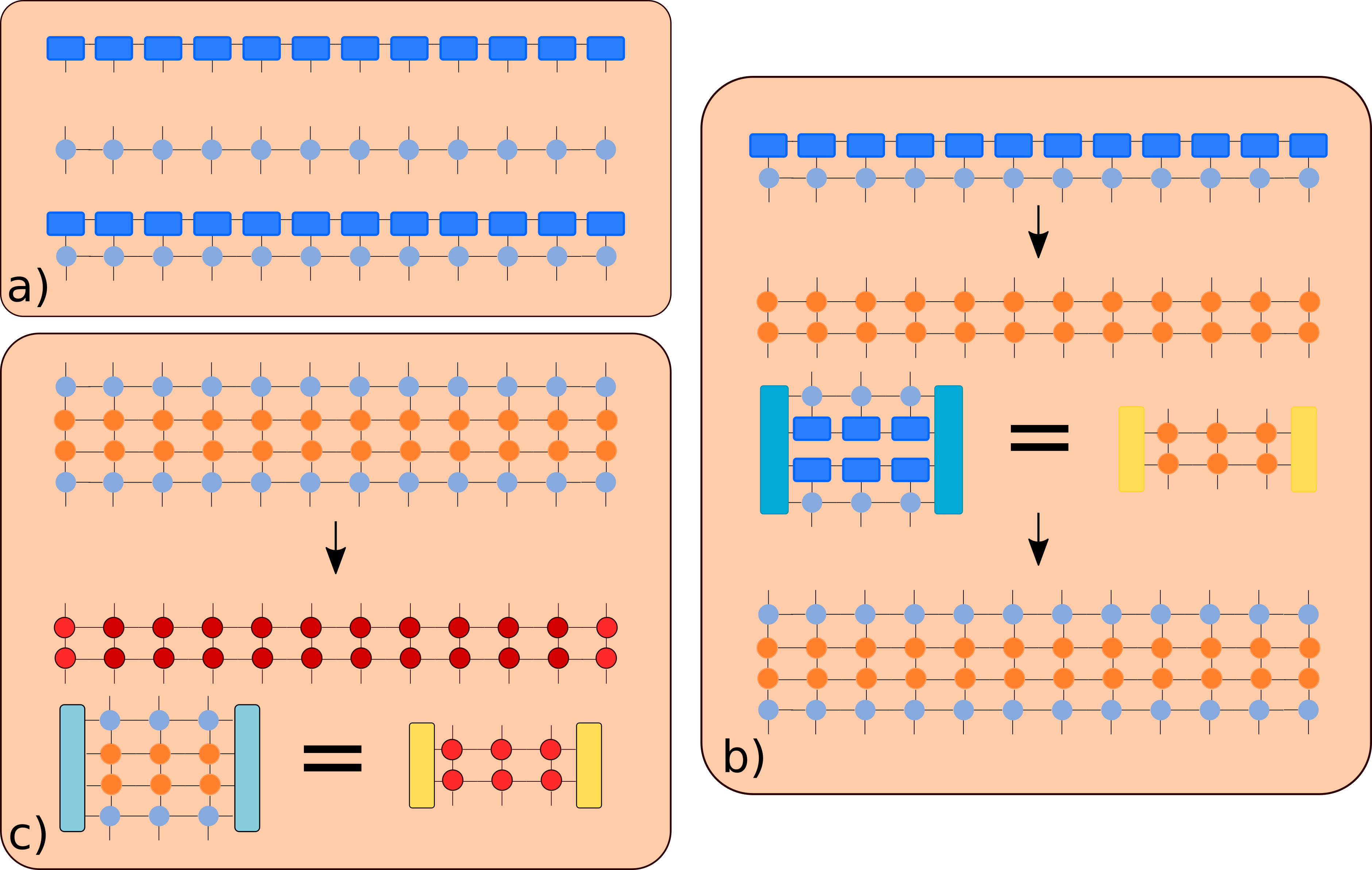}
 \caption{Tensor network scheme for the proposed algorithm. We use the standard tensor network notation where geometric shapes are tensors and their indices are represented by legs. Tensor contractions are represented by lines joining two different shapes. {\bf a)} An initial state that can be represented  by a tensor network is evolved for a short time. The initial state is represented as a blue MPS, the evolution is encoded by a pale-blue MPO. The evolved system is represented by the contraction of the MPS and the MPO. {\bf b)} The evolved state becomes highly entangled and thus we   approximate it with a mixed state represented here by an orange MPDO. The MPDO is obtained variationally by imposing that its reduced density matrices up to distance $m$ coincide exactly with those of the evolved state. Here  $m=3$. Once the best MPDO approximation of the evolved state is obtained, the system is evolved again for short times. {\bf c)} The subsequent dynamics increases again the  complexity beyond the  one we can deal with. We  repeat the approximation by using an MPDO with lower operator-Schmidt rank (in red in the figure). During the approximation we always force the  local indistinguishability of the approximate state and the evolved one, up to blocks of size $m$.
 \label{fig:scheme}}
\end{figure}

The $A$ are thus tensors with three indices, that are typically represented by  geometric shapes with three attached lines, one for each index (see panel a of Fig. \ref{fig:scheme}). When the indexes are contracted (such as in the matrix-matrix multiplications defining the MPS state), the corresponding lines are joined together. As a result, the graphical representation of an MPS state is made by several elementary three leg tensors (small blue boxes in Fig. \ref{fig:scheme})  connected by a line. In panel a) of  Fig.  \ref{fig:scheme} we illustrate an MPS state for  $N=12$ constituents.

The locality of the interactions allows also to approximate the operator $U(\delta t)$, for  $\delta t \ll 1$, at arbitrary order in $\delta t$ as a matrix product operator (MPO). A MPO  is an operator of the form $\hat{O}=\sum_{i_1 \cdots i_N,j_1 \cdots j_N} o^{i_1 \cdots i_N}_{j_1 \cdots j_N} \ket{i_1, \cdots i_N}\bra{j_1 \cdots j_N}$, and $ o^{i_1 \cdots i_N}_{j_1 \cdots j_N}= tr \left( B^{i_1}_{j_1}\cdot B^{i_2}_{j_2}\dots B^{i_{N-1}}_{j_{N-1}}\cdot B^{i_N}_{j_N}\right)$, where each $B^{i}_{j}$ is again a matrix. As a result the full $B$ are now tensors with four indices, and can be represented as geometric shapes with four attached lines. In panel a) of Fig \ref{fig:scheme} the $B$ are represented as pale-blue circles, and contracted to  represent the operator $U(\delta t)$ for a chain of $N=12$ constituents. 

The size of the $B^{i}_{j}$ matrices increases exponentially as we want to get higher order approximations to $U(\delta t)$. As a result, one typically sticks to relatively low-order approximations of  $U(\delta t)$  in $\delta t$  (e.g. 4th order) and carefully  reduces the $\delta t$ in order to get the required accuracy. We can thus assume that the matrices $B^{i}_{j}$ have a fixed rank $D_B$. 

We can perform a single step of time evolution (the TE part of the algorithm) in the tensor network notation by joining the appropriate lines of the MPO and the MPS in order to perform the required matrix vector multiplication. The TE  is  represented in the lower part of panel a) of Fig. \ref{fig:scheme}, where we compute $\ket{\psi(t_1)}=U(\delta t) \ket{\psi(0)}$ for $\ket{\psi(0)}$ expressed as a MPS and  $U(\delta t)$ encoded by a MPO.

The graphical representation makes explicit   that $\ket{\psi(t_1)}$ is again a MPS, with individual tensors given by $\tilde{A}^j=\sum_i A^i B_i^j$. The  bond dimension of $\tilde{A}^j$ has increased from $D$ to $D D_B$. This fact implies that the process cannot be iterated arbitrarily many times since the size of the matrices would increase exponentially  growing to $D D_B^n$ after $n$ iterations. 

We thus need to introduce an approximation step, that in the original framework is called the block decimation (BD). The idea introduced by Vidal in \cite{vidal2004a}, is to simplify 
$\ket{\psi(t_1)}$  by projecting the MPS matrices $\tilde{A}^j$ onto the subspace spanned just by  the most relevant  Schmidt vectors. These are the vectors  with the largest Schmidt weight.

If the state $\ket{\psi(t_1)}$ is slightly entangled, this projection can be done almost exactly and effectively reduces the size of the MPS matrices $\bar{A}^j$ to  $\bar{D} \ll  D D_B$. In these specific scenarios, one can thus iterate the procedure successfully several times and perform the evolution of the system for moderately long times. 

In a generic out-of-equilibrium scenario, however, the entanglement of a partition of the state grows linearly in time \cite{calabrese2005}. This means that the number of degrees of freedom required to describe a block of the system grows exponentially in time and the  BD truncation step of the standard TEBD algorithm produces a   $\bar{D} \simeq  D D_B$. The truncation step thus fails to reduce the computational cost of the dynamic and  the TEBD evolution remains exponentially expensive. 

In generic cases, the TEBD simulations are limited to short times since very soon the required computational resources exceed the available ones. Though we have presented the standard TEBD algorithm the same reasoning applies to any algorithm trying to perform the time-evolution of a state in a MPS form. 
 
Here we present  an alternative algorithm, that aims at using  a different approximation step to the BD of the TEBD and a different encoding of the state.  Similarly to the TEBD algorithm, our algorithm starts with an exact small step of time evolution $\delta t$, as illustrated 
%The scheme for such an algorithm is presented in Fig. \ref{fig:scheme}, where we illustrate the case of a one dimensional system using the standard tensor network notation. 
in panel a) of  Fig. \ref{fig:scheme}. We also assume that initially the state of   the system is  encoded in an MPS. We assume that during the short-time evolution generated by the MPO  encoding $U(\delta t)=\exp[-iH\delta t]$ the rank of the MPS increases to a value we can still safely encode. 

We want to be able to iterate arbitrarily many time the procedure in the cases were we know that the rank of the MPS would increase exponentially with the number of iterations. We thus need to trade the MPS representation for a different one that should make the algorithm  scalable to larger times. 

%We indeed know that (at least for large-enough times) the complexity of the evolved states is encoded in very non-local correlations, and the corresponding states are locally indistinguishable from simpler mixed states. This is indeed the essence of thermalization.  We thus i) compute the reduced density matrix of a region of $m$ spins, by tracing on the remaining $N-m$ spins. This is graphically represented in panel b of Fig. \ref{fig:scheme}, where the contraction of all the spins to the left and right of the $m=3$ region is represented by vertical rectangles and the tensor network contraction leading to the reduced density matrix of the three spins is explicitly shown on the left of the second line. We can now proceed to a Schmidt decomposition of the reduced density matrix, and a result express it as a matrix-product density operator, MPDO (shown in orange in the second line of panel b of Fig. \ref{fig:scheme}). The following step is ii) to  neglect the boundary terms in the TN network contraction giving rise to the reduced density matrix of the state, thus obtaining local tensors that if contracted appropriately  can be used to reconstruct a globally mixed state. That state by construction is locally indistinguishable from the the evolved pure state.  

We indeed know that (at least for large-enough times) the complexity of the evolved states is encoded in very non-local correlations, and the corresponding states are locally indistinguishable from simpler mixed states. This is indeed the essence of thermalization and suggests we should try to use mixed-states rather than pure ones. 

In order to find such mixed states, we first compute the reduced density matrix of a region of $m$ spins, by tracing out  the remaining $N-m$ spins. This is graphically represented in panel b) of Fig. \ref{fig:scheme}, where the contractions of all the spins to the left and right of the $m=3$ region are represented by vertical rectangles. The resulting reduced density matrix of three spins is encoded in the tensor network contraction shown on the left of the second line of panel b) of Fig. \ref{fig:scheme}. 

We proceed by performing a Schmidt decomposition of the reduced density matrix expressing the result as a matrix-product density operator, MPDO \cite{verstraete2004a,zwolak2004} (shown in orange in the second line of panel b of Fig. \ref{fig:scheme}).  Appealing to the translational invariance of the state  we neglect the boundary terms in the TN contraction of the reduced density matrix of the state, and use the local MPDO tensor in order to build a trial mixed state. That state is mixed, but not yet locally indistinguishable from the evolved pure state. We thus start the variational optimisation that will ensure the local indistinguishability.

Morally thus, the algorithm consists in trading the evolved state (the first line of the panel b) for  a mixed state $\rho(0)$ encoded by a MPDO represented in orange in the second line of the panel. The tensors defining the MPDO are variationally optimised, by starting from the tensors obtained from the Schmidt decomposition of the reduced density matrix of a region as we just described.

The cost function of the optimisation  (a measure of the distance between the original reduced density  matrix and the one obtained from the mixed state)  forces  the resulting mixed state to become locally indistinguishable from the evolved state. This condition  is encoded in the equality in the third line of the panel. There the vertical rectangles (blue on the left and yellow on the right) encode the contraction of the tensor network originating from tracing the constituents outside the region we are interested in and can be easily computed with standard tensor network techniques.

We stress once more that the choice of a local cost function  for the variational optimisation of the tensors, together with trading pure states for mixed states, constitute the main differences with respect to the MPS based algorithm. If most of the complexity of the evolved state comes from correlations at large distances (as it does in simple quenches), the MPDO bond dimension (that encodes the correlations at short distances) can be reduced at much smaller value than the bond dimension required to describe the original state \footnote{A simple example is the state of four spin $1/2$ constituents $ \ket{S(24)S(13)}$ where the $S(ij)=\frac{1}{\sqrt{2}}\left(\ket{0_i0_j}+\ket{1_i1_j}\right)$. This state written as an MPS $c^{i_1,i_2,i_3,i_4}=A{i_1} A{i_2}A{i_3}A{i_4}$ requires $A$ with bond dimension $2$. However, the reduced density matrix $\rho(23)$ is the tensor product $\rho(2)\rho(3)$ and thus has operator Schmidt rank 1.}.
 
The maximum MPDO  Schmidt rank is also fixed by the available computational resources as in any other tensor network algorithm.  In addition to the MPDO Schmidt rank,  the size of the region over which the MPDO and the original state are indistinguishable constitutes a new refinement parameter $m$. When $m\to N$, where $N$ is the system size, we are morally very close to MPS based algorithms. Here we thus focus on the opposite limit, where we expect our algorithm to strongly deviate from the standard ones. We  want to characterise the algorithm at fixed $m \ll N$. This is the reason why in Fig.   \ref{fig:scheme} we have presented the case of $m=3$.

As for the standard TEBD algorithms, once the optimal MPDO is found, we iterate the procedure, as shown in panel c) of Fig. \ref{fig:scheme}. Here we evolve the system for a certain extra time $\rho(t) = U^{\dagger}(t)\rho(0) U(t)$, shown in the first line of the panel,  and then we truncate  its operator Schmidt rank to the pre-established maximal value obtaining the MPDO represented in red in the second line of the panel. The truncation is performed always variationally, without  affecting the local correlations as a consequence of the equality in the third raw of the panel. 

Summarising, the main idea is  to transform the entanglement present in the initial state into mixture, something that is natural in the context of thermalisation at large times. In order to perform this transformation gradually, starting already at relatively short times, we need to force that  the mixed states we use are locally  indistinguishable from the evolved state. 
%$As a result if such mixed state exists, using the known reultsSince all mixed state obey the area-law and are thus representable as MPDO \cite{wolf2008}, 

Our algorithm thus implements Jayne's principle \cite{jaynes1957,jaynes1957a}, we are designing an approximate dynamics where at each step the entropy grows due to the fact that some of the generated entanglement is transformed into mixture. At the same time,  the ``relevant''  conserved quantities  (i.e. conserved quantities built out of local densities)  are protected from the approximations and thus kept constant. We thus expect that, as a consequence of the robustness of the equilibration, the process will  equilibrate to a state locally  indistinguishable from the DE. Furthermore, in the process we avoid diagonalising the full Hamiltonian \cite{perarnau-llobet2016}, something unavoidable if trying to directly construct the DE.

While trading entanglement with mixture is clearly possible at very long  time (this is,  in the end,  the essence of thermalisation), we need to understand if this is feasible at the early stages of the evolution. We thus need to understand if mixed states that are locally indistinguishable from the states produced in early stages of the evolution exist, how to construct them, and the effects they induce once used at a given time as the starting point for the subsequent evolution.

Since these are general questions that do not  depend on the tensor-network formulation of the problem, we  address them in the context of free Fermionic models. This allows us to separate any methodological difficulty  from the physical effects that such approximation will produce. Such methodological difficulties include for example: the choice of the norm to use in order to force the equalities in Fig. \ref{fig:scheme} that define the MPDO; assuming that such MPDO exists how to construct an initial guess for it that can be variationally improved; how to design a TN algorithm guaranteed to converge to the optimal MPDO starting from the initial guess; what is the extra effect of the finite operator Schmidt rank.

%The separation between the short-range correlations, which are accurately  reproduced, and the long-range correlations, which are ignored,  constitute the natural refinement  parameter of the algorithm.  The refinement parameter  affects both the precision of the computation and the resources required to perform it. When the refinement parameter approaches the size of the system, the dynamics becomes exact.

The relevant states of  free Fermionic systems are Gaussian (see Appendix \ref{Appendix:Gaussian}). As a specific example we will consider the transverse field Ising model
\begin{equation}
H(\theta) =  -\sin(\theta)\sum_{i=0}^{N-1}\sigma_{i}^{x}\sigma_{i+1}^{x}-\cos(\theta) \sum_{i=0}^{N-1} \sigma_{i}^{z},
\label{eq:TFI-Hamiltonian}
\end{equation}
that can be re-expressed as a free-Fermionic Hamiltonian
\begin{align}
H(\theta) =  &-\sin(\theta)\sum_{i=0}^{N-1}\left[ a_i^{\dagger}a_{i+1} -a_ia_{i+1}^{\dagger} +a_{i}^{\dagger}a_{i+1}^{\dagger}-a_{i}a_{i+1} \right]+ \nonumber \\
&-\cos(\theta)\sum_{i=0}^{N-1}\left[a_{i}^{\dagger}a_{i}-a_{i}a^{\dagger}_{i} \right].
\label{eq:TFI-Fermionic form}
\end{align}
Because of the Wick theorem a generic Gaussian state is fully characterised by the two point correlation functions which can be organised in the so called symbol or correlation matrix \cite{greplova2013,kraus,schultz1964}

\begin{equation}
\Lambda_{i,j}=\langle \vec{\alpha}_i \vec{\alpha}^{\dagger}_j\rangle,
\label{eq:symbol-matrix}
\end{equation}
where  $\vec{\alpha}^{\dagger}=\left(a_1,a_2,\dots,a_N,a_1^{\dagger},a_2^{\dagger},\dots,a_{N}^{\dagger}  \right)$ is the collection of annihilation and creations operators for every site.

We benchmark our algorithm by studying the out-of-equilibrium evolution generated by a sudden quench of the Hamiltonian. We start from  the ground state of the Hamiltonian  \eqref{eq:TFI-Fermionic form} for a given $\theta_0$, $H(\theta_0)$   encoded in the correlation matrix $\Lambda_0$ and we let it evolve with an  Hamiltonian having the same structure, but defined with a different value of  $\theta$, $H(\theta)$. 

The out-of-equilibrium evolution of free-Fermionic systems can be computed exactly providing the ideal setting for benchmarking our approximate algorithm as we explain in detail in \ref{Appendix:Gaussian}, \ref{Appendix: G-evolution}.

% In particular we want be able to  approximate locally the relaxed state using as little information as possible about the state at each step of the evolution, aiming to obtain an efficient algorithm to predict locally the state emerging at long times after the quench.

%The approximate algorithm should,  i) contain  a refinement parameter $m$ allowing to tune the precision. For small values of $m$ we  only obtain very rough results. They should systematically improve and converge to the exact ones as $m$ is increased; ii) preserve those conserved quantities expressed as the sum of local enough densities (for example, the energy or the number of particles). The privileged role of these conserved quantities in the context of the out-of-equilibrium evolution has been widely discussed in the literature \cite{fagotti2013}.

The steps of the algorithm outlined in Fig. \ref{fig:scheme} can be implemented directly at the level of the correlation matrices. In particular, the truncation step corresponds to defining a truncated matrix $T_m(\Lambda)$, with $m\in \left[0,\floor{\frac{N}{2}}\right]$,  obtained from $\Lambda$ by setting all the matrix elements corresponding to correlations at distances $d>m$ to zero. For every finite-size system made by  $N$ constituents,   as  $m$ grows to  $m=\floor{\frac{N}{2}}$,   $T_m(\Lambda)=\Lambda$ and  thus the approximation becomes exact.
 $T_m(\Lambda)$ indeed preserves all the reduced density matrices consisting of $m+1$ sites and, as a result, all the expectation values of local operators with support on $m+1$ consecutive sites thus implementing exactly the equality in the panels b) and c) of  \ref{fig:scheme}.

As an example, in the quenches discussed here, $m\geq1$ is enough to conserve the energy of the system since the total energy for a transverse field Ising Hamiltonian \eqref{eq:TFI-Fermionic form} is the sum of operators with support on only $2$ consecutive sites. For a generic Gaussian Fermionic Hamiltonian sum of operators with support on at most $l$ neighbouring sites, the conservation of the energy is enforced  by choosing $m\geq(l-1)$.

At last,  the truncation maps translation-invariant states to translation-invariant states, thus preserving translation invariance for every choice of $m$.
We can thus build an approximate time evolution algorithm by approximating  $\Lambda$ with $T_m(\Lambda)$ at every step of the  evolution. In the specific case of the free Fermions, the algorithm presented in Fig. \ref{fig:scheme} translates in the following pseudo-code,

\begin{algorithm}[H]
  \caption{Truncated time evolution of precision $m$}
  \label{EPSA}
   \begin{algorithmic}[1]
 \Procedure{Trunc-Evolv}{$\Lambda,N_s, \delta t,m$}%
 			\State $t\coloneqq 0$
            \While{$t<N_s$}
                \State $\Lambda\coloneqq T_m(\Lambda)$ \Comment{Truncation step}
                \State $\Lambda \coloneqq Evolve(\Lambda, H(\theta_1),\delta t)$ \Comment{Evolution step}
                \State $t \coloneqq t+1$
            \EndWhile\label{euclidendwhile}
            \State \textbf{return} $\Lambda$
        \EndProcedure.
   \end{algorithmic}
\label{Agorithm}
\end{algorithm}

The evolution time step is performed via exact diagonalisation on the space of the symbol matrices that scales linearly with the dimension of the system. Details on the evolutions are reported in appendix \ref{Appendix: G-evolution}.

While the above algorithm could give the illusion that mixed states sharing the same local reduced density matrices with a given state could always be found for Gaussian states, this is not the case.  Even though the truncation step preserves the local reduced density matrices of the system it does change the global state. We note, for example, that changing from a finite number  to $0$ the out-of-diagonal elements of a matrix, in general modifies its eigenvalues. The actual change can only be found by  diagonalising the matrix before and after the truncation. As a result, the approximation that we are performing by zeroing the correlation at distances larger than $m$  can in principle spoil the positivity of the state.

If the above truncation spoils the positivity of the state, this means that we would not be able to find a state in the Hilbert space represented by the correlation matrix we are using. This does not exclude that there could be another mixed state that is still locally indistinguishable from the state we want to approximate, but at least tells us that this state is possibly hard to identify. 
We will monitor this specific aspect of the  algorithm in the numerical analysis. 

Before  moving to the numerical results we note that the loss of information due to the truncation of the correlations makes the approximate dynamics not unitary. This is somehow expected.  We are indeed trying to obtain  a good approximation of the Gaussian diagonal ensemble $\rho_{GDE(H)}$ whose knowledge, in general, is not sufficient to recover the initial state of the quench. 

As a word of caution, our exact results  refer both to the full out-of-equilibrium dynamics of the system and to the Gaussian state $\rho_{GDE}$  (Gaussian diagonal ensemble \cite{j.surace}) built from the symbol matrix  $\Lambda_{DE}$ \eqref{eq:symbol-matrix} of the DE. In general this state can be different from the actual diagonal ensemble $\rho_{DE}$,  in the specific case we consider here, the two states are locally indistinguishable. With a slight abuse of notation we will thus refer to  the $\rho_{GDE}$ as $\rho_{DE}$ \footnote{In general the possible discrepancies between $\rho_{GDE}$ and $\rho_{DE}$ have already been  largely studied in  the literature, e.g. $\rho_{GDE}$ has been characterised numerically in  \cite{rigol2007} (where it is called \emph{fully constrained thermodynamic ensemble}) and through the recent analytical calculations presented in  \cite{murthy2018,gluza2018} (where $\rho_{GDE}$ is called \emph{Gaussian Generalised Gibbs Ensemble}), and we thus refer the interested reader to those publications.}.

\section{Numerical results}
We study the dynamics for the quench  ${\theta: \frac{\pi}{4}+0.1\rightarrow\frac{\pi}{4}+0.3}$ of the Hamiltonian $\eqref{eq:TFI-Fermionic form}$, a quench inside the ordered phase. The results obtained for different quenches both  in the disordered phase and across the phase transition are qualitatively similar  as can be checked in our appendix \ref{Appendix:different-quenches}.  We compare different system sizes, $N=1500$ sites (that we use as the thermodynamic limit) and smaller system sizes. For $N=1500$ we compute the exact post-quench dynamics for very long times,  up to the corresponding recurrence time $T_R\propto 1500$. These are the results we consider exact and we use as a benchmark.

As an intermediate  system we consider $N=200$ sites. For this system we compute the dynamics for all the admitted values of the parameter $m$ up to the time $T_R$. In all scenarios we use time-steps of length $\delta t=0.25$. The smallest system we consider contains $N=41$ sites since its exact evolution requires similar computational resources to the one required by our approximation of the larger systems. 

%Since we are dealing with free Fermionic systems the time evolution after the quench can be computed exactly and we can thus compare the results of the approximated algorithm with those obtained during the exact evolution.
%As shown in \ref{Agorithm}, each step of the approximated algorithm consists of two transformations of the matrix $\Lambda$: first one applies to $\Lambda$ the non-unitary transormation $T_{M}(\Lambda)$  and a subsequently one computes the exact unitary time evolution of $\Lambda$  for an interval $\delta t$.
%
%The operation $T_{m}(\Lambda)$, that we call truncation, sets to $0$ all the elements of $\Lambda$ that correspond to correlations at distance greater than $m$. We call $m$ the \textit{refinement parameter}.
We indicate with $n=a^{\dagger}a$ the single site occupation operator (the choice of the specific site is irrelevant since the system is translationally invariant). We represent as $\rho_N(t,m)$ the density matrix of the system of $N$ sites evolved to time $t$ with the truncated algorithm with a value of the refinement parameter $m$ (when $m$ is not explicitly specified we are referring to the state evolved exactly). The expectation value of an operator $O$ on a state $\rho$ is indicated as $\langle O \rangle_{\rho} = Tr\left[\rho O\right]$.

\begin{figure}[!hbt]
 \includegraphics[width=\columnwidth]{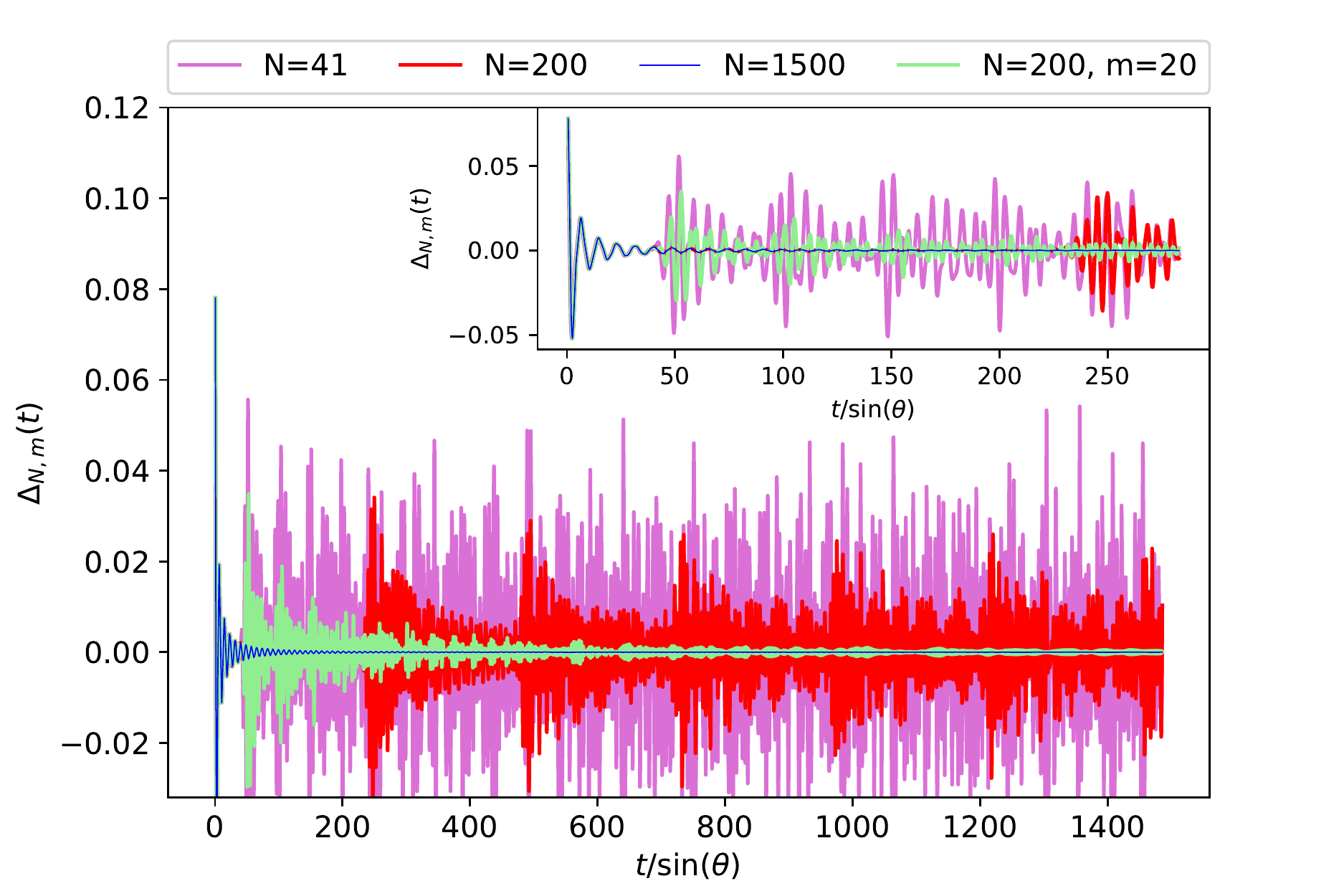}
 \caption{Time evolution of the quantity $\Delta_{N,m}(t) \coloneqq \langle n \rangle_{\rho_{N}(t,m)}-\langle n \rangle_{GDE}$ for different values of the parameters $N$ and $m$. In the inset a zoom on the first part of the dynamics where the recurrence effects for two exacts evolution and the approximation error deriving from the truncations are visible. In the main picture we plot the evolution for long times. The truncated dynamic converges towards the GDE more slowly than the exact one.}
 \label{fig:figure1}
\end{figure}

In figure \ref{fig:figure1} we focus  on characterising the dynamics of $n$. All other local operators behave similarly, as we show in the appendix \ref{Appendix:observable-independence}  where we focus on the evolution of the two sites reduced density  matrices that encode the expectation value of  arbitrary operators defined on two consecutive sites. 

We plot the deviation of the dynamics of the single site occupation $n$ from its equilibration value given by the GDE, for several systems evolved both exactly and approximately
% {\bf FORMULA HERE}
\begin{equation}
\Delta_{N,m}(t) \coloneqq \langle n \rangle_{\rho_{N}(t,m)}-\langle n \rangle_{GDE}.
\end{equation}
The equilibrium value of $\langle n \rangle_{\rho_{1500}(t)}$ tends algebraically towards the value predicted by the GDE. This is a consequence of the general results about  the equilibration rate (measured as the damping of the of the envelope of the oscillations in figure \ref{fig:figure1}) discussed in \cite{murthy2018,gluza2018}. The specific operator we are considering converges towards the GDE as a power law, proportional to  $ (t^{-\frac{3}{2}})$. This rate of convergence is in perfect agreement  with the predictions for the thermodynamic limit contained in \cite{bucciantini2014,calabrese2012}, confirming that  the size  of the system is large enough to be considered infinite.

In  the exact dynamics of the smaller systems  ($N=40$ and $N=200$) we can clearly identify the recurrence effects by the rebirth of large oscillations at later times. It is worth noting that before the recurrence effects become evident, the dynamics of the local observable is the same for all the sizes of the systems we have considered. This is easier to observe in the inset of  \ref{fig:figure1} representing a zoom of the main plot for short times, $\frac{t}{\sin(\theta)}\in[0,280]$. There we can also appreciate that, as expected, the recurrence time is proportional  to the size of the system (in the thermodynamic limit the proportionality constant is the maximum group velocity of the pseudo particles \cite{calabrese2005,calabrese2011,calabrese2012a,calabrese2012}).

%oIn the intervals $\frac{t}{\sin(\theta)}\in [0,\sim 800]$  and $\frac{t}{\sin(\theta)}\in [0,\sim 170]$ the dynamics of $\langle n \rangle_{\rho_{200}}(t)$ and $\langle n \rangle_{\rho_{1500}}(t)$ overlap.
In the plot we also present the truncated dynamics, $\langle n \rangle_{\rho_{200}(t,20)}$. The choice of $m=20$ implies that the truncation step always preserves the reduced density matrices of all the  sub-systems of $m+1=21$ consecutive sites and for each site it preserves all the correlation with $2m+1=41$ sites. In order to understand the difference between our truncation scheme and the evolution of a small finite system we also compare the results of the approximate dynamics with those of the exact dynamics for a small system of exactly $N=41$ sites.

%We start by characterising  the late-time behaviour of local observables that should be well approximated (at least on average) by their value in the DE \cite{rigol2009}.
%The occupation of a single site $n\coloneqq a^{\dagger}a$ (all sites are equivalent as a consequence of translational invariance) extracted from the approximate dynamics is  $\langle n(t) \rangle_m\coloneqq Tr\left[\rho(t,m)n \right]$ while its exact value is  $\langle n (t) \rangle\coloneqq Tr\left[\rho(t)n \right]$ and its  value in the diagonal ensemble $\langle n\rangle_{DE}$.
%%%%%%%%%%%%%%%%%%%%%%%%%%%%%%%%%%%%%FIGURE 1%%%%%%%%%%%%%%%%%%%%

%%%% CAMBIA NOME ENVELOPE PERCHE' Δ è già usato nella figura 2  %%%%
%%%%%%%%%%%%%%%%%%%%%%%%%%%%%%%%%%%%%%%%%%%%%%%%%%%%%%%%%

%The corresponding results are presented in the top panel of Figure \ref{fig:SV1} where we plot the  evolution after the quench  of  $\langle n(t) \rangle_m$ versus  $\langle n(t) \rangle$ together with  $\langle n\rangle_{DE}$.
%During the exact evolution the expectation value of $n$ relaxes after moderately long times (of the order $t/sin(\theta) = 60$ in Figure \ref{fig:SV1}) to an almost stationary value, well approximated by $\langle n\rangle_{DE}$. If the system were infinite this would be the end of the out-of-equilibrium evolution, but due to the finite size of the system, after times of the order of $N/v$ (where $v$ is the maximum group velocity of the pseudo particles \cite{calabrese2005,calabrese2011,calabrese2012,calabrese2012a}), the pseudo particles have already travelled  through the full system causing return effects and the dynamics to restart.

As expected, the truncation does not affect the dynamics at short time.
Since the initial state is the ground state of a gapped Hamiltonian its correlation functions decay exponentially with the distance \cite{hastings2007,wolf2008,eisert2013}. During the initial steps of the evolution $\langle n(t) \rangle_{\rho_{200}(t,20)}$ is extremely close  to $\langle n(t)\rangle_{\rho_{200}(t)}$.
When the correlations spread to distances larger than $m$, the approximation induced by the truncation becomes evident.

It is instructive to compare $\rho_{41}(t)$ with $\rho_{200}(t,20)$. In the exact evolution of the small system, the correlated regions spread-apart but eventually, due to the periodic boundary conditions, meet again. They meet  when each correlation front has travelled through half of the system, in this case $21$ sites. When they meet, the recurrence effects they produce disturb the equilibration process of the system. 

In contrast, in the truncated evolution the correlations are destroyed once spread further than  $21$ sites.  Erasing such correlations completely changes the dynamics that deviates both from the one of the small periodic system and from that of the larger (practically infinitely large) system.

In order to quantify  these modifications we compute the truncated dynamics far beyond the recurrence time of the system with $N=2m+1=41$. In the main figure we observe how $\langle n \rangle_{\rho_{200}(t,20)}$ slowly converges towards a value close to the one predicted by the GDE. We call $e(m)$ the value towards which each truncated evolution with parameter $m$ converges. We need to both characterise how $e(m)$ depends on $m$ and how fast the truncated dynamics converges to $e(m)$.

\begin{figure}[H]
	\includegraphics[width=\columnwidth]{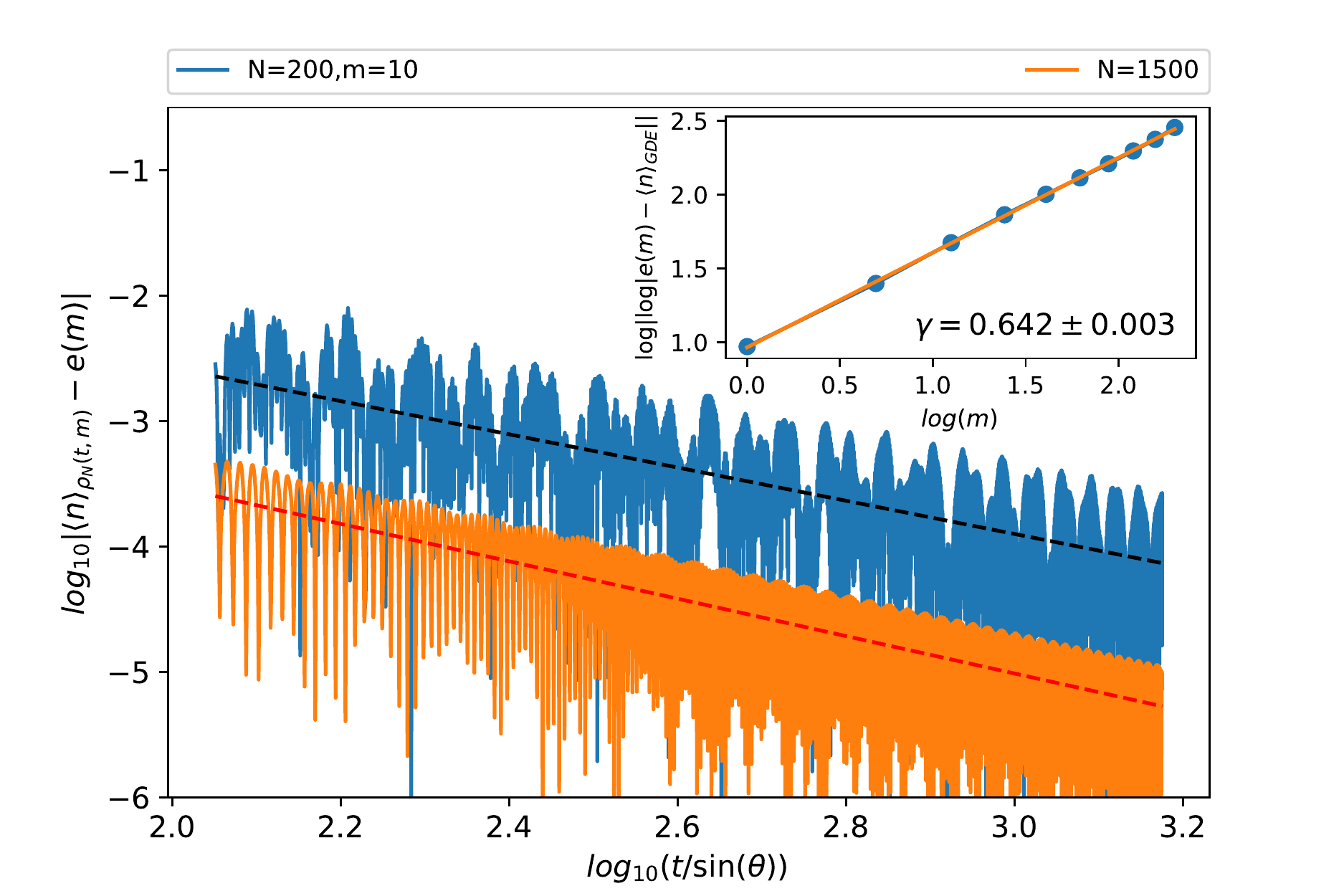}
	\caption{{\bf{(Main)}}  Logarithmic difference between $\langle n\rangle_{\rho_{N}(t,m)}$ and the expected equilibration value at a given $m$, $e(m)$, as a function of the logarithm of time. The exact dynamics converges algebraically to  $\langle n \rangle_{GDE}$. The approximate dynamics converges algebraically to $e(10)$. The two dotted lines are linear fits to the data of the dynamics. The quantity $\langle n \rangle_{\rho_{N}(t)}$ converges to $\langle n \rangle_{GDE}$ as $t^{-\frac{3}{2}}$ for $t<T_{R}$, where $T_R$ is the recurrence time for the given $N$. We qualitatively see that the truncated dynamics converges towards $e(10)$ with a similar trend.
	{\bf(Inset)} Here we address the dependence of $e(m)$ on $m$. We plot the log-log difference between the equilibration values $e(m)$ and $\langle n \rangle_{GDE}$ as a function of  $\log(m)$. The linear dependence suggests that $e(m)$ converges towards $\langle n \rangle_{GDE}$ as $(e(m)-\langle n \rangle_{GDE}) \sim e^{-m^{\gamma}}$. 
	Our best fit gives an estimate  $\gamma=0.642 \pm 0.003$.	
	}
	\label{fig:figure2}
\end{figure}

Both analyses are performed in figure \ref{fig:figure2} where we study the trend of convergence of the truncated dynamics towards its equilibration value $e(m)$. We furthermore study the dependence of the equilibration value $e(m)$ from $m$.

Considering the time interval during which the approximation induced by the truncated dynamic is manifest we study the evolution of the quantity $ \log|\langle n\rangle_{\rho_{N}(t,m)}-e(m)|$.

We perform a linear fit on both the exact and truncated dynamics. For the exact dynamics we considered $e(\infty)=\langle n \rangle_{GDE}$ as equilibration value, in this case the slope of the fitting line is $\sim \frac{3}{2}$ as expected. For the truncated evolution we notice that the rate of convergence towards $e(m)$ is qualitatively similar to the one of the exact dynamics.

At this stage it is important to notice that  the exact dynamics of a finite size system  will always present recurrences, while the truncated dynamics equilibrates to $e(m)$ for every value of $m$. The dependence of $e(m)$ on $m$ is addressed in the inset of figure \ref{fig:figure2}. The data suggest a dependence 
\begin{equation}
e(m) = e^{-m^{\gamma}}+\langle n \rangle_{GDE}
\label{eq:convergence-ansatz}
\end{equation}
with a value of $\gamma = 0.642 \pm 0.003$ extracted by performing a best-fit of equation \ref{eq:convergence-ansatz} to our data.

While we have focused here on a specific local observable and a specific quench protocol, similar results are obtained for all other local observables and quench protocols as shown in the appendices \ref{Appendix:different-quenches} and \ref{Appendix:observable-independence}.

%%%%%%%%%%%%%%%%%%%%%%%%%%%%%%%%%%%%%FIGURE 3%%%%%%%%%%%%%%%%%%%%
\begin{figure}[H]
 \includegraphics[width=0.95\linewidth]{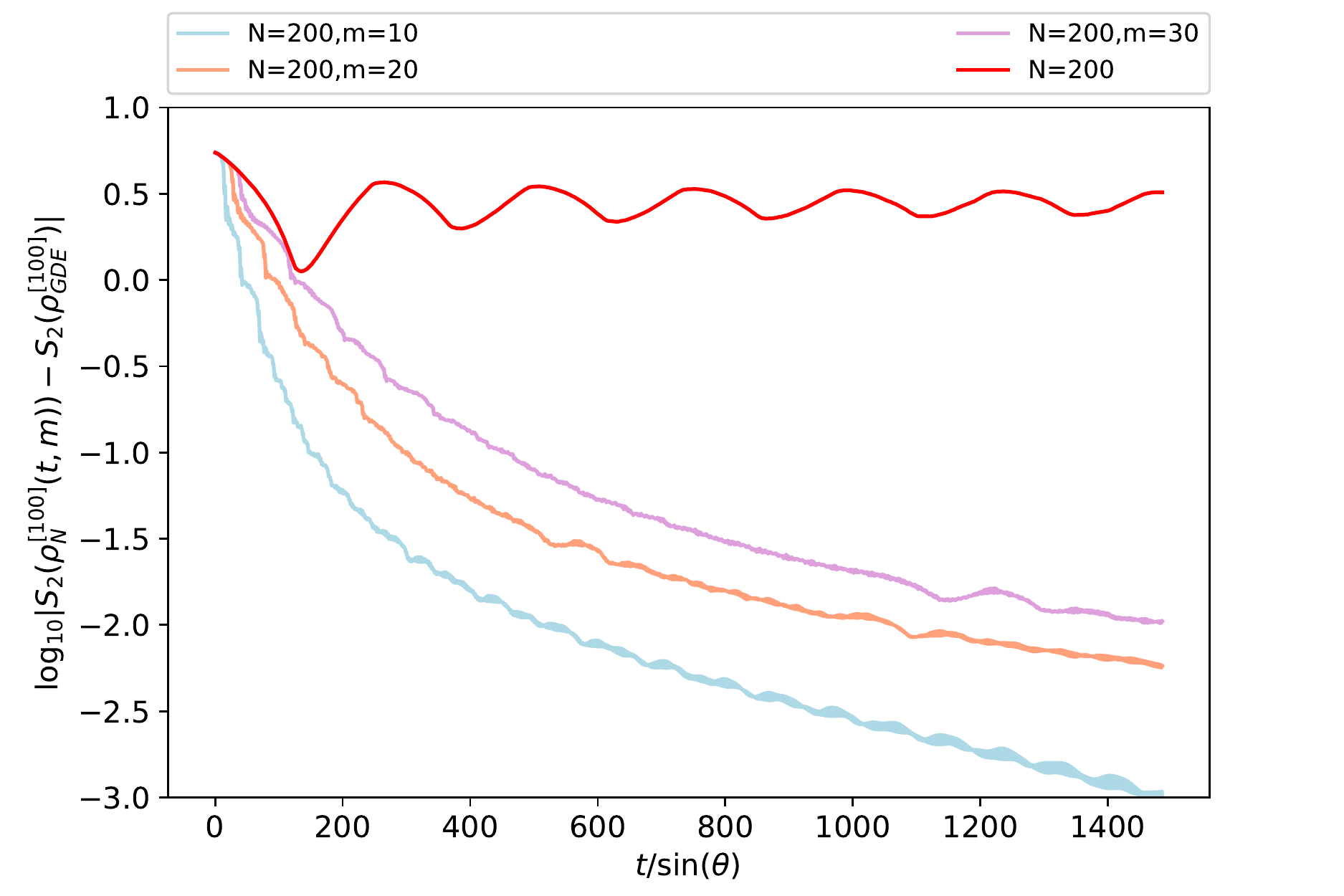}
 \caption{Time evolution of $|S_2\left(\rho^{[100]}_{N}(t,m) \right)-S_2(\rho^{[100]}_{GDE})|$ in logarithmic scale for different values of the parameters $N$ and $m$. In the exact dynamics the recurrence effects are evident from the oscillatory behaviour of the dynamics. In the truncated dynamics, correlations are not allowed to return into the partition, therefore, once spread outside, they are lost forever. We see that indeed the entropy always increases getting closer to $S_2(\rho^{[100]}_{GDE})$ as $m$ increases. 
% 	{\bf where are the lines for N=1200? I would eliminate 8 unless you think it says something relevant, like the fact that it is not a physical state}.
}
 \label{fig:figure3}
\end{figure}
%%%%%%%%%%%%%%%%%%%%%%%%%%%%%%%%%%%%%%%%%%%%%%%%%%%%%%%%%

%The truncated evolution thus provides a good approximation  of the local properties of the GDE.

We now turn to characterise the state of the system we obtain after the truncation steps.
We first check that, as already mentioned, the purity of $\rho_{200}(t,m)$ decreases during the evolution encoding the fact that the state becomes mixed. The truncation step of the algorithm adds mixedness to the global system, while at the same time (for $m>1$) conserving local densities, in perfect accordance with Jayne's principle.

In order to confirm this, we consider the reduced state of a block of $100$ consecutive sites described by the reduced density matrix $\rho^{[100]}_{N}(t,m)=Tr_{[101,\dots,N]}\left[\rho_{N}(t,m)\right]$.

We study the evolution of the second Renyi entropy, defined for generic density matrix $\rho$ as ${S}_2=- \log(Tr\left[ {\rho}^2\right])$ in figure \ref{fig:figure3} where we plot the time evolution of $|S_2\left(\rho^{[100]}_{N}(t,m) \right)-S_2(\rho^{[100]}_{GDE})|$, in logarithmic scale, for different values of the parameters $N$ and  $m$.

For the exact dynamics with $N=200$, $S_2$  grows close to the  value of $S_2(\rho^{[100]}_{GDE})$ before starting to decrease as a result of the recurrence. It then oscillates, with a frequency fixed by the size of the system.

In the truncated dynamics, correlations are not allowed to return into the partition, therefore, once spread outside, they are lost forever. We see that, indeed, the entropy always increases getting closer to $S_2(\rho^{[100]}_{GDE})$ as $m$ increases.

%The entropy can even exceed that of  the GDE, that contains a larger amount of constrains than the one we conserve by only considering short-range correlations as seen for example  in the case of $m=8$ in figure \ref{fig:figure3}.  . This is confirmed by the fact that has we increase the range of correlations we conserve the entropy converges from above towards its GDE value.  

% {\bf CAN THIS HAPPENS FOR PHYSICAL STATES, does it mean that the state is not physical isnt GDE the one with maximal entropy? Yes it can happen, the GDE is the one with maximal entropy and with ALL the conserved quantity conserved. The truncated GDE, conserving a smaller number of quantities has more freedom, thus can maximise the entropy even more.}

%Similarly to the case of the local occupation, the second Renyi entropies reaches a steady value closer and closer to the the GDE as we increase $m$
% {\bf NOT TRUE, does it means again the the best choice of m is the one for which the entropy grows closer to the GDE value?. The Value get closer and closer similiarly to how the value of the truncated GGE goes closer and closer to the one of the GGE as one add more conserved quantities.}.

%%%%%%%%%%%%%%%%%%%%%%%%%%%%%%%%%%%%%FIGURE 4%%%%%%%%%%%%%%%%%%%%
\begin{figure}[H]
 \includegraphics[width=\linewidth]{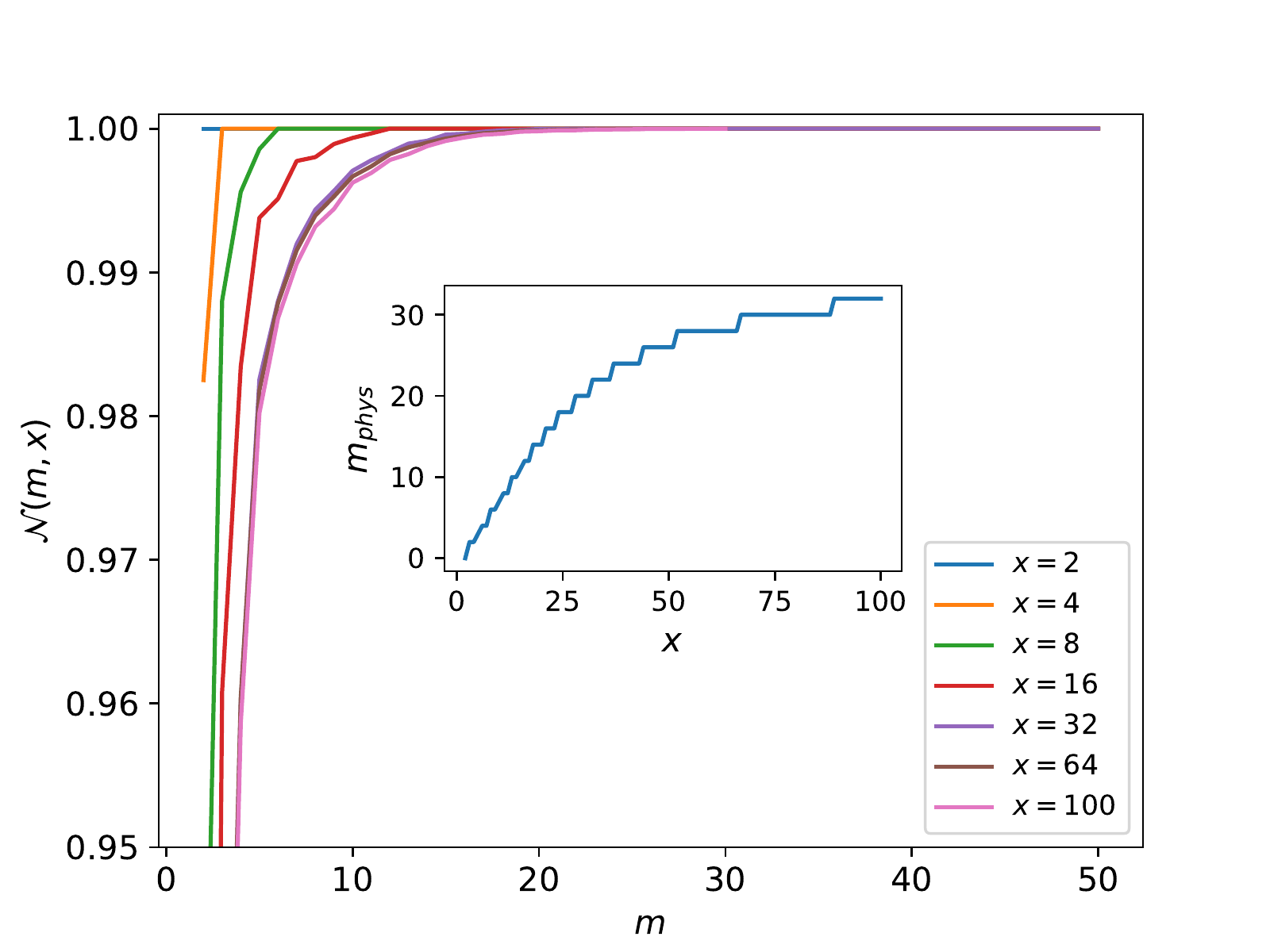}
 \caption{{\bf{(Main)}} The value of $\mathcal{N}(m,x)$ versus $m$ is plotted for different dimensions $x$ of the reduced matrix $\overline{\rho}(m)_x$. When $\mathcal{N}(m,x)=1$ then $\overline{\rho}(m)_x$ is physical.
 {\bf(Inset)} Minimum value of the precision $m_{phys}$ for a specific $x$ such that $\overline{\rho}(m)_x$ is physical.
 It is remarkable that already for moderate values of $m$ the approximate state is physical over a large range of distances.
}
 \label{fig:means}
\end{figure}
%%%%%%%%%%%%%%%%%%%%%%%%%%%%%%%%%%%%%%%%%%%%%%%%%%%%%%%%%

It is important to check if the states generated by our approximate dynamics are physical. A priori, the symbol matrix obtained after the truncation step is not guaranteed to represent a legitimate state of the free-Fermionic Hilbert space.
A matrix is a density matrix only if it is  positive semi-definite.  Checking if a matrix is positive requires diagonalising it, and, since the size of the matrix increases exponentially with the size of the system, this becomes unfeasible  already for very small systems. In the case of tensor networks, the task is shown to be NP-hard in the dimension of the system \cite{kliesch2014}.\\
Conversely, the symbol matrix $\Lambda$ of a free-Fermionic system represents a physical state if  its eigenvalues are in the interval $[0,1]$ \cite{kraus}. 
%This point has been completely ignored in the literature due to the exponential complexity of the task. Here again free Fermions help addressing this issue with only polynomial resources.

In order to check if on average the physicality is conserved, we consider the density matrix  $\bar{\rho}(m)$ associated to the time average of the symbol matrices for each $m$,
\begin{equation}
\overline{\Lambda}(m) = \frac{1}{N_s}\sum_{t=1}^{N_s}\Lambda(t \cdot\delta t,m),
\label{eq:average matrix}
\end{equation}
where $N_s$ is the number of steps considered for the evolution.
This check requires diagonalising the symbol matrix, an operation that scales polynomially with the system size. The eigenvalues of $\Lambda$ appear in couples $(\lambda_i,1-\lambda_i)$.  $\Lambda$ is non-physical if it contains at least one eigenvalue smaller than $0$ (hence its partner will be larger than $1$).

We thus define the quantity $\mathcal{N}(m,x)$, that is $1$ minus the sum of the negative eigenvalues of $\Lambda(m)_{x}$,
\begin{equation}
\mathcal{N}(x,m) = 1-\frac{||\Lambda(m)_x||_{l=1}-1}{2}=1-\sum_{i}\frac{|\lambda_i|-\lambda_i}{2},
\end{equation}
where $\Lambda(m)_x$ is the symbol matrix of $Tr_{[1,\dots,N-x]}\left[\overline{\rho}(m)\right]=\overline{\rho}(m)_x$ and  $\{\lambda_i\}_{i}$ are the eigenvalues of $\Lambda(m)_x$.\\
$\mathcal{N}(m,x)=1$  for a physical $\overline{\rho}(m)_x$, since as mentioned all the eigenvalue are in this case positive in the interval between $0$ and $1$.  $\mathcal{N}(m,x)$  decreases as the absolute value of the sum of negative eigenvalues of $\Lambda(m)_x$ increases.

In Figure \ref{fig:means} we plot $\mathcal{N}(m,x)$ versus $m$ for different values of $x$.  We can see that for large enough  $m$ the reduced density matrices of $\overline{\rho}(m)$ become physical for every chosen size $x$. In the inset, we plot the minimum value of precision $m_{phys}$ required for $\overline{\rho}(m)_x$ to be physical for every choice of its size  $x$.

This fact should be related to the finite correlation length present in the GDE. In order to describe correctly the expectation value of a local operator we just need to embed the local system into a larger system whose size exceeds the correlation length of the desired state (see e.g. \cite{hernandez-santana2015,kliesch2014a,depasquale2016,garcia-saez2009,ferraro2012}).

\section{Conclusion}
We have identified some robust aspects of the out-of-equilibrium dynamics encoded in the late-time expectation value of local operators after the quench. By exploiting this robustness we have designed an approximate algorithm that allows to predict the relaxed values of local operators with limited computational resources.

The key idea underlying the design of our algorithm is to protect from the approximations the relevant conserved quantities. These are defined as the conserved quantities built out of local densities. The degree of locality of such conserved quantities naturally acts as the refinement parameter of the algorithm allowing to increase the precision of the results by increasing the computational cost.

We have benchmarked the algorithm in the case of free-Fermionic systems where the algorithm can be implemented very easily at the level of the correlation matrices. Here the approximation required by the algorithm corresponds to discarding those elements of the correlation matrices \eqref{eq:symbol-matrix} that are at distance larger than a certain cut-off distance  $m$ from the diagonals.

We have observed that for modest values of $m$ the results are in good agreement with the exact ones. Furthermore, their precision improves exponentially as we increase the computational resources.

%In particular we have identified the lower value of the refinement parameter that guarantees the physicality of the approximate states generated by the algorithm.

%We have observed that, in most cases, for moderate values of the refinement parameter $m$, our  algorithm provides a good local approximation to the DE, by practically generating a dynamics that monotonically increases the entropy while exactly protecting the local conserved quantities. 

The next step is to check if the same picture holds in the presence of strong interactions both in the integrable and non-integrable scenarios, by implementing the generic version of the algorithm using tensor networks along the lines of the scheme presented in Fig. \ref{fig:scheme}.
It would also be interesting to compare and relate this approach with the  existing complementary one proposed in  \cite{leviatan2017,white2018,wurtz2018,vonkeyserlingk2018,leviatan2017,caux2013,nardis2015,caux2016}.

We acknowledge the discussion on the topic with Frank Verstraete, Marie Carmen Banuls, Frank Pollman. JS was supported by the doctoral training partnership (DTP 2016-2017) of the University of Strathclyde. While this paper was under review several  proposals have appeared that design and benchmark alternative algorithms for simulating the long-time dynamics of the many-body systems efficiently  \cite{rams2019,krumnow2019,cao2017}.

\section{Appendix}
\subsection{Fermionic Gaussian systems}
\label{Appendix:Gaussian}
The generic quadratic (or Gaussian) Fermionic Hamiltonian on $N$ sites can be written as
\begin{equation}
H = \frac{1}{2}	\sum	_{i,j=1}^{N-1}\left[A_{i,j}a_i^{\dagger}a_j -\overline{A}_{i,j}a_ia_j^{\dagger}+B_{i,j}a_ia_j-\overline{B}_{i,j}a_i^{\dagger}a_{j}^{\dagger}\right]
\label{eq:Gen-quad-Hamiltonian}
\end{equation}
where $A_{i,j},B_{i,j}\in \mathbb{C}$, $A=A^{\dagger}$, $B^T=-B$ and the annihilation and creation operators $a_i$ and $a_i^\dagger$ obey the anticommutation relations $\{a^{\dagger}_i,a_j\}=\delta_{i,j}$, $\{a^{\dagger}_i,a^{\dagger}_j\}=\{a_i,a_j\}=0$.\\
In the main text we restricted to the single parameter Hamiltonian $H(\theta)$ \eqref{eq:TFI-Fermionic form} defined by the matrices
\begin{align}
A(\theta)_{i,j} &=-f_{N,i,j}\sin(\theta)(\delta_{i+1,j}+\delta_{i,j+1})-2\cos(\theta)\delta_{i,i},\nonumber \\
B(\theta)_{i,j} &=f_{N,i,j}\sin(\theta)(\delta_{i+1,j}-\delta_{i,j+1}),
\label{parameters}
\end{align}
with the term
\begin{equation}
f_{N,i,j}=
\begin{cases}
-(-1)^{(N)}, & \text{if}\ i \vee j=N \\
1, & \text{otherwise},
\end{cases}
\end{equation}
necessary for the antisymmetrisation with periodic boundary condition.\\
This one  parameter Hamiltonian is the mapping to Fermions of the $1D$ transverse field Ising model on $N$ sites and periodic boundary condition
\begin{equation}
H(\theta) =  -\sin(\theta)\sum_{i=0}^{N-1}\sigma_{i}^{x}\sigma_{i+1}^{x}-\cos(\theta) \sum_{i=0}^{N-1} \sigma_{i}^{z},
\end{equation}
where  $(\sigma^x_i, \sigma^x_i, \sigma^z_i)$ are the Pauli matrices on site $i$, $\theta \in [0,\frac{\pi}{2}]$ and $N$ is the number of sites\cite{dutta2012}.\\

A Fermionic Gaussian state is a ground or thermal state of a quadratic Fermionic Hamiltonian. Using Wick's theorem it is possible to show that Gaussian states are completely characterised by the collection of their $2-$points correlators
\begin{align}
\Lambda^{TL}_{i,j} &= Tr\left[\rho a_i^{\dagger} a_j \right] \\
\Lambda^{TR}_{i,j} &= Tr\left[\rho a_i^{\dagger} a_j^{\dagger} \right],
\end{align}
where the correlators $Tr\left[\rho a^{\dagger}_i a_j \right]$ and $Tr\left[\rho a^{\dagger}_i a^{\dagger}_j\right] $ are said to be a correlators at distance $d_{i,j}$,  where $d_{i,j} \equiv \min(|i-j|, |N-(i-j)|)$.
We have that $\Lambda^{TL}$ is Hermitian and  $\Lambda^{TR}$ is skew-symmetric. The $2-$points correlators can be arranged in the block matrix
\begin{align*}
\Lambda=
\left[
\begin{array}{c|c}
\Lambda^{TL} & \Lambda^{TR} \\
\hline
-{\Lambda^{TR}}^{*} & \mathbb{I}-{\Lambda^{TL}}^{T}
\end{array}
\right],
\end{align*}
called the symbol matrix \cite{greplova2013}.
The symbol matrix is Hermitian and for any admissible symbol matrix of a physical Fermionic system, the eigenvalues have to be in the interval $[0,1]$.\\
The $2-$points correlators of translational invariant states depend only on the distance, thus the elements $\Lambda^{TL}_{i,j}$ simplify to elements of the circulant matrices $\Lambda^{TL}_{d_{i,j}}$ which depend only on the distance between the indices.\\
The space of Gaussian states is closed under evolution induced by quadratic Hamiltonians, thus, if we start with a Gaussian state and evolve it with a Hamiltonian of the form \eqref{eq:Gen-quad-Hamiltonian} the knowledge of $\Lambda$ at any time would completely characterise the state of the system.\\
The reduced state over a set of sites $S$ is still a Gaussian state and it is characterised by the symbol matrix of all the $2-$points correlators with support on $S$.

\subsection{Gaussian evolution of Fermionic Gaussian  states}
\label{Appendix: G-evolution}
A generic Gaussian Fermionic Hamiltonian $H$ of the form \eqref{eq:Gen-quad-Hamiltonian} can always be brought in the form
\begin{equation}
H=\frac{1}{2}\sum_{k=0}^{N-1}\epsilon_k\left(b^{\dagger}_k b_k-b_k b^{\dagger}_k \right) \end{equation}
for a specific set of Fermionic annihilation and creation operators $b_k,b^{\dagger}_k$.

The time evolution of the operators $b_k,b^{\dagger}_k$ with the Hamiltonian $H$ is easily computed as:
\begin{align}
b_k(t) = e^{-i \epsilon_k}t b_i, \nonumber \\
b^{\dagger}_k(t) = e^{i \epsilon_k} t b^{\dagger}_k.
\end{align}

Given a symbol matrix $\Lambda$ expressed on the basis of the Fermionic annihilation and creation operators $a_i,a^{\dagger}_i$, there exist 
%a suitable transformation that maps  the modes $a_i,a^{\dagger}_i$ to $b_k,b^{\dagger}_k$ and a corresponding  
unitary operation that allows one to express $\Lambda$ on the basis of $b_k,b^{\dagger}_k$. In term of these operators the submatrices of $\Lambda$ can be written as:
\begin{align}
\Lambda^{TL}_{k,l} = Tr\left[\rho b^{\dagger}_k b_l \right], \nonumber \\
\Lambda^{TR}_{k,l} = Tr\left[\rho b^{\dagger}_k b^{\dagger}_l \right].
\end{align}
Using this representation, one is able to compute the time evolution of the correlation matrix  elements as
\begin{align}
\Lambda^{TL}_{k,l}(t) = e^{i(\epsilon_k-\epsilon_l)} \Lambda^{TL}_{k,l}, \nonumber \\
\Lambda^{TR}_{k,l}(t) = e^{i(\epsilon_k+\epsilon_l)} \Lambda^{TR}_{k,l}.
\end{align}
One can then return to the basis of the annihilation and creation operators $a_i,a^{\dagger}_i$.
\subsection{Numerical results for different quenches}
\label{Appendix:different-quenches}
\begin{figure}[H]
	\includegraphics[width=\columnwidth]{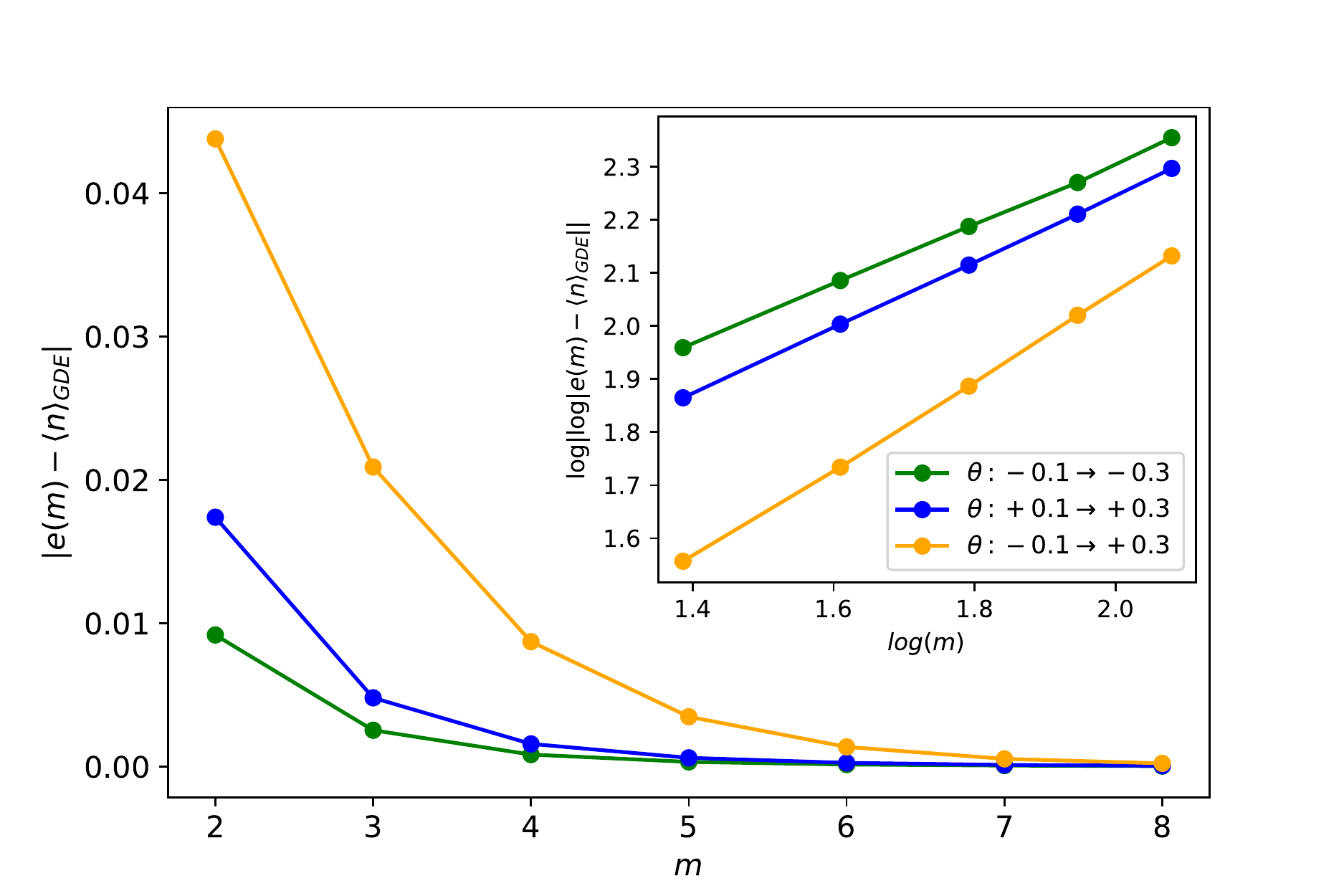}
	\caption{{\bf Main } The difference $|e(m)-\langle n \rangle_{GDE}|$, where $e(m)$ is the equilibrium value of the local observable $n$ for the truncated evolution of parameter $m$ in the corresponding quench (different quenches correspond to different colours) and $\langle n \rangle_{GDE}$ is the value of $n$ computed on the corresponding GDE. {\bf Inset} We plot the same data of the main figure, with a suitable scale, in order to check the validity of the ansatz \eqref{eq:convergence-ansatz}.}
	\label{fig:different-quenches}
\end{figure}

We checked the algorithm for different quenches of the Hamiltonian \eqref{eq:TFI-Fermionic form}. 
In the main text we analysed the dynamic for a quench above the critical point ($\theta=\frac{\pi}{4}$). We study the dynamics for quenches below the critical point (${\theta: \frac{\pi}{4}-0.1\rightarrow\frac{\pi}{4}-0.3}$) and across the critical point (${\theta: \frac{\pi}{4}-0.1\rightarrow\frac{\pi}{4}+0.3}$).
In figure \ref{fig:different-quenches} we plot the difference $|e(m)-\langle n \rangle_{GDE}|$, where $e(m)$ is the equilibrium value of the local observable $n$ for the truncated evolution of parameter $m$ in the corresponding quench (different quenches correspond to different colours). The linear trend in the inset confirms the validity of the ansatz \eqref{eq:convergence-ansatz}.

\subsection{Observables independent local convergence}
\label{Appendix:observable-independence}
We checked the local convergence of the $2$-sites reduced density matrices $\rho^{[2]}_{N}(m,t) \coloneqq Tr_{[3,\dots,N]}\left[\rho_{N}(m,t)  \right]$ towards the $2$-sites reduced density matrix of the GDE $\rho^{[2]}_{GDE}  \coloneqq Tr_{[3,\dots,N]}\left[\rho_{GDE}  \right]$.

\begin{figure}[H]
	\includegraphics[width=\columnwidth]{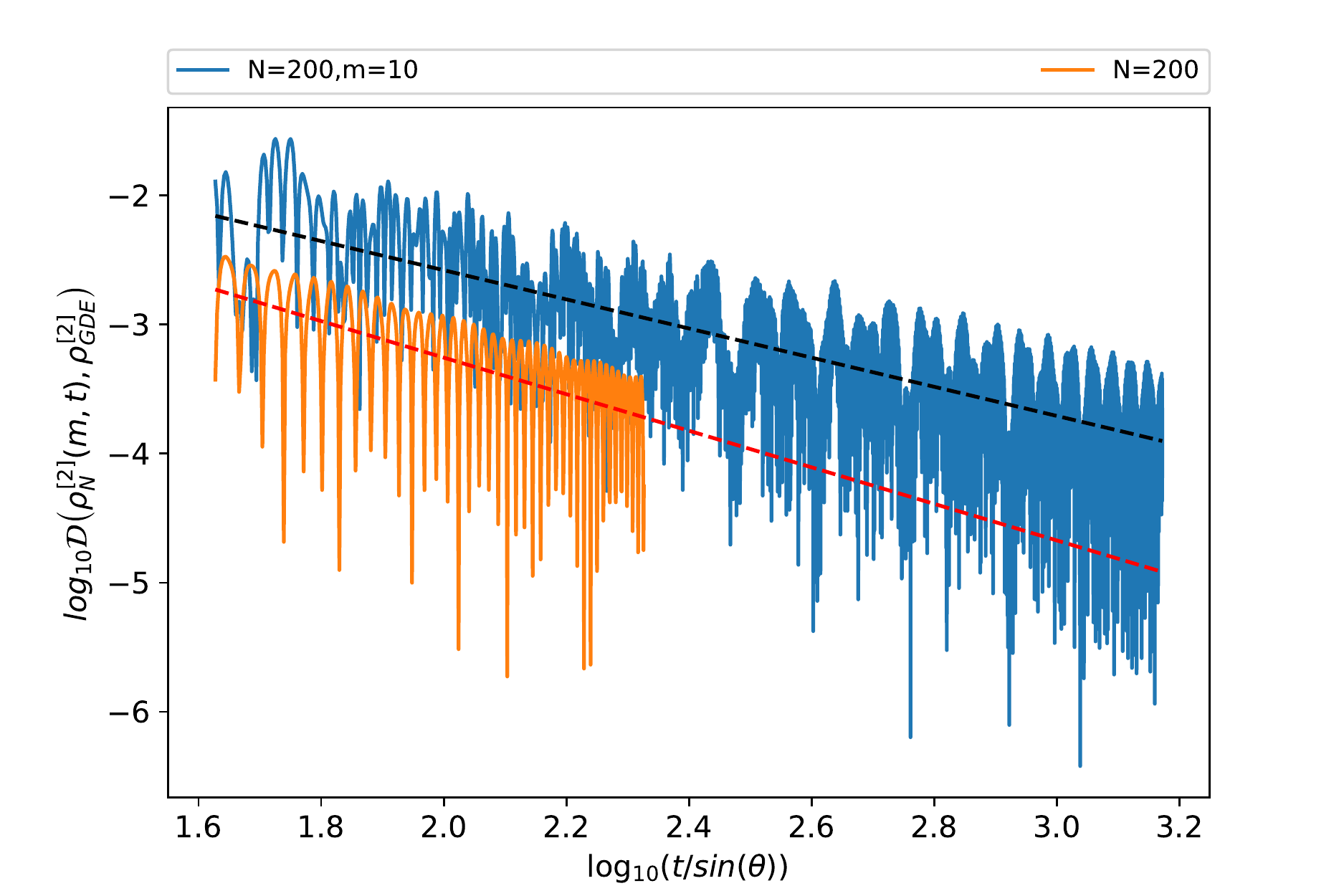}
	\caption{Time evolution of the logarithm of the trace distance $ \mathcal{D}\left(\rho^{[2]}_{N}(m,t),\rho^{[2]}_{GDE} \right)$. Both the exact dynamics as the truncated dynamics locally converge towards the GDE.
	}
	\label{fig:trace-distance}
\end{figure}

In figure \ref{fig:trace-distance} we plot the time evolution of the logarithm of the trace distance 
\begin{equation}
\mathcal{D}\left(\rho^{[2]}_{N}(m,t),\rho^{[2]}_{GDE} \right) = \frac{1}{2} Tr\left[\left|\rho^{[2]}_{N}(m,t)-\rho^{[2]}_{GDE} \right| \right].
\end{equation}
The trace distance is a measure of the maximum probability of distinguishing between two states with an optimal measurement.

The two dotted lines are the linear fits for the two distances. 

The $2 \times 2$ reduced system, both for the exact as for the  truncated dynamics, converge towards the $2 \times 2$ reduced system of the GDE with a similar trend as the one observed in the observable dependent scenario of figure \ref{fig:figure2}.

%%%%%%%%%%%%%%%%%%%%%%%%%%%%%%%%%%%%%%%%%%%%
%From figures \ref{fig:Ameans} and \ref{fig:Bmeans} we note that the minimum precision $m_{phys}$ for the physicality of the whole system is again proportional to the correlation length after the quench. For shorter correlation length a small value of the precision $m$ is sufficient in order to obtain a globally physical states that locally well approximate the diagonal ensemble.
\bibliography{LOED-PRB.bib}
%\bibliography{/home/jacopo/Dropbox/Mendeley/LOED.bib}
\bibliographystyle{apsrev4-1}

\end{document}